\documentclass[a4paper,10pt]{article}
\usepackage{stmaryrd}
\usepackage{amsfonts}
\usepackage{bbm}
\usepackage{amscd}
\usepackage{mathrsfs}
\usepackage{latexsym,amssymb,amsmath,amscd,amscd,amsthm,amsxtra,xypic}
\usepackage[dvips]{graphicx}
\usepackage[utf8]{inputenc}
\usepackage[T1]{fontenc}
\usepackage{lmodern}
\usepackage{amssymb}
\usepackage[all]{xy}
\usepackage{nicefrac,mathtools,enumitem}
\usepackage{microtype}
\usepackage{lmodern}

\newcommand {\emptycomment}[1]{}
\newtheorem{thm}{Theorem}[section]
\newtheorem{lem}[thm]{Lemma}
\newtheorem{cor}[thm]{Corollary}
\newtheorem{pro}[thm]{Proposition}

\newtheorem{rmk}[thm]{Remark}
\newtheorem{defi}[thm]{Definition}

\newcommand{\lon }{\,\rightarrow\,}
\newcommand{\be }{\begin{equation}}
\newcommand{\ee }{\end{equation}}

\newcommand{\pf}{\noindent{\bf Proof.}\ }

\newcommand{\g}{\frkg}

\newcommand{\huaA}{\mathcal{A}}

\newcommand{\huaV}{\mathcal{V}}

\newcommand{\frkg}{\mathfrak g}
\newcommand{\frkh}{\mathfrak h}

\newcommand{\frkk}{\mathfrak k}
\newcommand{\frkl}{\mathfrak l}

\newcommand{\frkG}{\mathfrak G}

\def\qed{\hfill ~\vrule height6pt width6pt depth0pt}

\newcommand{\half}{\frac{1}{2}}

\newcommand{\Courant}[1]{\left\llbracket  #1\right\rrbracket_{\beta} }

\newcommand{\br}[1]{   [ \cdot,    \cdot  ]   }
\newcommand{\id}{\mathbbm{i}}
\newcommand{\idd}{\mathrm{id}}

\newcommand{\dM}{\mathrm{d}}

\newcommand{\Ad}{\mathrm{Ad}}

\newcommand{\gl}{\mathfrak {gl}}

\newcommand{\ad}{\mathrm{ad}}

\newcommand{\sgn}{\mathrm{sgn}}

\begin{document}
\title{
{On hom-Lie algebras
} }
\author{Yunhe Sheng and Zhen Xiong \\
Department of Mathematics, Jilin University,
 Changchun 130012,  China
\\\vspace{3mm}
email: shengyh@jlu.edu.cn,~xiongzhen13@mails.jlu.edu.cn }
\date{}
\footnotetext{{\it{Keyword}: hom-Lie algebras, representations, omni-hom-Lie algebras, hom-Lie $2$-algebras }}
\footnotetext{{\it{MSC}}: 17B99, 55U15}

\maketitle

\begin{abstract}
  In this paper, first we show that $(\g,[\cdot,\cdot],\alpha)$ is a hom-Lie algebra if and only if $(\Lambda \g^*,\alpha^*,d)$ is an $(\alpha^*,\alpha^*)$-differential graded commutative algebra. Then, we revisit representations of hom-Lie algebras, and show that there are a series of coboundary operators. We also introduce the notion of an omni-hom-Lie algebra associated to a vector space  and an invertible linear map. We show that regular hom-Lie algebra structures on a vector space can be characterized by Dirac structures in the corresponding omni-hom-Lie algebra. The underlying algebraic structure of the omni-hom-Lie algebra is a hom-Leibniz algebra, or a hom-Lie 2-algebra.
\end{abstract}


\section{Introduction}

The notion of hom-Lie algebras was introduced by Hartwig, Larsson,
and Silvestrov in \cite{HLS} as part of a study of deformations of
the Witt and the Virasoro algebras. In a hom-Lie algebra, the Jacobi
identity is twisted by a linear map, called the hom-Jacobi identity.
Some $q$-deformations of the Witt and the Virasoro algebras have the
structure of a hom-Lie algebra \cite{HLS,hu}. Because of close relation
to discrete and deformed vector fields and differential calculus
\cite{HLS,LD1,LD2},   more people pay special attention to this algebraic structure. In particular, hom-Lie algebras on semisimple Lie algebras are studied in \cite{jin1}; Quadratic hom-Lie algebras are studied in \cite{BM}; Representation theory, cohomology and homology theory are systematically studied in \cite{AEM,chengsu,homlie1, Yao2}; Bialgebra theory and hom-(Classical) Yang-Baxter Equation are studied in \cite{MS3,shengbai,Yao1,YaoCYBE}; The notion of a hom-Lie 2-algebra, which is a categorification of a hom-Lie algebra, is introduced in \cite{homlie2}, in which the relation with hom-left-symmetric algebras  \cite{MS2} and  symplectic hom-Lie algebras are studied.

In this paper, first we study the structure on the exterior algebra $\Lambda \g^*$ associated to a hom-Lie algebra $(\g,[\cdot,\cdot],\alpha)$. We generalize  the result ``A Lie algebra structure on a vector space $\frkk$ is equivalent to a differential graded commutative algebra structure on the exterior algebra $\Lambda \frkk^*$'' to the case of a hom-Lie algebra. Then, we go on to study the structure on $\Lambda\g^*\otimes V$ associated to a representation of a hom-Lie algebra $(\g,[\cdot,\cdot],\alpha)$ on a vector space $V$ with respect to $\beta\in\gl(V)$. We find that the results about representations of Lie algebras can not be generalized directly to the case of hom-Lie algebras due to the restriction of hom-cochains. Only part of result can be obtained. Furthermore, motivated by Jin's work, we construct a hom-Lie algebra structure on $\gl(V)$ using any invertible $\beta\in GL(V)$. Consequently, some representations can be described by a morphism between hom-Lie algebras. In particular, the  adjoint representation $\ad$ of a regular hom-Lie algebra $(\g,[\cdot,\cdot],\alpha)$ is a morphism from $(\g,[\cdot,\cdot],\alpha)$ to $(\gl(\g),[\cdot,\cdot]_\alpha,\Ad_\alpha)$.  Sometimes, it is helpful to describe a morphism by the structure on its graph. For example, $f:\huaA_1\longrightarrow\huaA_2$ is an algebra morphism if and only if the graph of $f$ is a subalgebra of $\huaA_1\oplus \huaA_2$, where $\huaA_1, \huaA_2$ are associative algebras. Now, it is natural to ask what is the structure on the graph of the adjoint representation $\ad:\g\longrightarrow\gl(\g)$? To solve this question, we introduce the notion of an omni-hom-Lie algebra $\gl(V)\oplus_\beta V$ associated to a vector space $V$ and an invertible map $\beta\in\ GL(V)$. The graph of $\ad$ is a Dirac structure. However, the algebraic structure on  $\gl(V)\oplus_\beta V$ is not a hom-Lie algebra. It is a hom-Leibniz algebra ( \cite{LD2, MS2}, if nonskewsymmetric bracket is used), or a hom-Lie 2-algebra (if a skew-symmetric bracket is used).

The paper is organized as follows. In Section 2, we recall some
necessary background knowledge, including hom-Lie algebras and their representations, and hom-Lie 2-algebras. In Section 3, we introduce the notion of a $(\sigma,\tau)$-differential graded commutative algebra, and show that $(\frkg,[\cdot,\cdot],\alpha)$ is hom-Lie algebra if and only if
$(\Lambda \g^*,\alpha^*,d)$ is an $(\alpha^*,\alpha^*)$-differential graded commutative algebra (Theorem \ref{thm:dgca}). In Section 4, first we construct a hom-Lie algebra   $(\gl(V),[\cdot,\cdot]_\beta,\Ad_\beta)$ using an invertible map $\beta\in GL(V)$. A representation of $(\frkg,[\cdot,\cdot],\alpha)$ on $V$ with respect to $\beta\in GL(V)$ can be described by a morphism from $(\frkg,[\cdot,\cdot],\alpha)$ to $(\gl(V),[\cdot,\cdot]_\beta,\Ad_\beta)$ (Theorem \ref{thm:rephom}). Then, we introduce a series of coboundary operators $d^s,~s=0,1,2,\ldots,$ associated to a representation of a hom-Lie algebra. The properties of $d^s$ are given in Theorem \ref{thm:rep}. In Section 5, we introduce the notion of an omni-hom-Lie algebra, which we denote by $(\frkg\frkl(V)\oplus V,\delta_\beta,\{\cdot,\cdot\}_\beta,\langle\cdot,\cdot\rangle)$. We show that hom-Lie algebra structures on $V$ can be characterized by Dirac structures on the corresponding omni-hom-Lie algebra (Theorem \ref{thm1}). The bracket $\{\cdot,\cdot\}_\beta$ in an omni-hom-Lie algebra is not skewsymmetric, and $(\frkg\frkl(V)\oplus V,\{\cdot,\cdot\}_\beta,\delta_\beta)$ is a hom-Leibniz algebra. We also introduce the bracket $\Courant{\cdot,\cdot}$, which is the skewsymmetrization of the bracket $\{\cdot,\cdot\}_\beta$, and show that the algebraic structure behind $\Courant{\cdot,\cdot}$ is a hom-Lie 2-algebra (Theorem \ref{thm:homLie2}).

\section{Preliminaries}

\subsection{Hom-Lie algebras and their representations}

The notion of a hom-Lie algebra was introduced in \cite{HLS}, see also \cite{BM,MS2} for more information.

Let $\huaA$ be an associative algebra, and $\sigma$ and $\tau$ denote two   algebra endomorphisms on $\huaA$. A {\bf $(\sigma,\tau)$-derivation} on $\huaA$ is a linear map $D:\huaA\longrightarrow \huaA$ such that
\begin{equation}
  D(ab)=D(a)\tau(b)+\sigma(a)D(b),\quad \forall~a,b\in\huaA.
\end{equation}
In particular, a {\bf $\sigma$-derivation} on $\huaA$ is a $(\sigma,\idd)$-derivation.

In \cite{HLS}, the authors showed that there is a hom-Lie algebra structure on the set of $\sigma$-derivations of an associative algebra $\huaA$.

\begin{defi}
\begin{itemize}
\item[\rm(1)]
  A hom-Lie algebra is a triple $(\frkg,\br ,,\alpha)$ consisting of a
  vector space $\frkg$, a skewsymmetric bilinear map (bracket) $\br,:\wedge^2\frkg\longrightarrow
  \frkg$ and an algebra endomorphism $\alpha:\frkg\lon\frkg$ satisfying the following hom-Jacobi
  identity:
  \begin{equation}
   [\alpha(x),[y,z]]+[\alpha(y),[z,x]]+[\alpha(z),[x,y]]=0,\quad\forall
x,y,z\in\frkg.
  \end{equation}

 A hom-Lie algebra is called a regular hom-Lie algebra if $\alpha$ is
an algebra  automorphism.

 \item[\rm(2)] A sub-vector space $\frkh\subset\frkg$ is a hom-Lie sub-algebra of $(\frkg,\br ,,\alpha)$ if
 $\alpha(\frkh)\subset\frkh$ and
  $\frkh$ is closed under the bracket operation $\br,$, i.e. for all $ x,y\in\frkh$,
  $[x,y] \in\frkh.  $
  \item[\rm(3)] A morphism from the  hom-Lie algebra
$(\frkg,[\cdot,\cdot]_{\frkg},\alpha)$ to the hom-Lie algebra
$(\frkh,[\cdot,\cdot]_{\frkh},\gamma)$ is a linear map
$\psi:\frkg\longrightarrow\frkh$ such that
$\psi([x,y]_{\frkg})=[\psi(x),\psi(y)]_{\frkh}$ and
$\psi\circ \alpha =\gamma\circ \psi$.
  \end{itemize}
\end{defi}

Hom-Leibniz algebras are generalizations of hom-Lie algebras, see \cite{LD2, MS2} for more details.
\begin{defi}A hom-Leibniz algebra is a triple $(\frkg,\br ,,\alpha)$ consisting of a
  vector space $\frkg$, a   bilinear map (bracket) $\br,:\wedge^2\frkg\longrightarrow
  \frkg$ and an algebra endomorphism $\alpha:\frkg\lon\frkg$ satisfying the following hom-Leibniz rule
  \begin{equation}
   [\alpha(x),[y,z]]=[[x,y],\alpha(z)]+[\alpha(y),[x,z]],\quad\forall
x,y,z\in\frkg.
  \end{equation}
\end{defi}

 Representation and cohomology theories of hom-Lie algebra are
systematically introduced in \cite{AEM,homlie1}. See \cite{Yao2} for homology theories of hom-algebras. See \cite{chengsu} for more details about representation   theories for hom-Leibniz algebras.

\begin{defi}
  A representation of the hom-Lie algebra $(\frkg,\br,,\alpha)$ on
  a vector space $V$ with respect to $\beta\in\gl(V)$ is a linear map
  $\rho:\frkg\longrightarrow \gl(V)$, such that for all
  $x,y\in\frkg$, the following equalities are satisfied:
  \begin{eqnarray}
 \label{representation1} \rho (\alpha(x))\circ \beta&=&\beta\circ \rho (x);\\\label{representation2}
    \rho([x,y] )\circ
    \beta&=&\rho (\alpha(x))\circ\rho (y)-\rho (\alpha(y))\circ\rho (x).
  \end{eqnarray}
\end{defi}

\subsection{Hom-Lie 2-algebras}

The notion of a hom-Lie 2-algebra is introduced in \cite{homlie2}, which is a categorification of a hom-Lie algebras. A hom-Lie 2-algebra is equivalent to a 2-term $HL_\infty$-algebra.
\begin{defi}\label{defi:2hl}
  A hom-Lie $2$-algebra $\huaV$ consists of the following data:
\begin{itemize}
\item[$\bullet$] a complex of vector spaces $V_1\stackrel{\dM}{\longrightarrow}V_0,$

\item[$\bullet$] bilinear maps $l_2:V_i\times V_j\longrightarrow
V_{i+j}$,

\item[$\bullet$] two linear transformations $\phi_0\in\gl(V_0)$ and $\phi_1\in\gl(V_1)$ satisfying $\phi_0\circ\dM=\dM\circ\phi_1$,
\item[$\bullet$] a skewsymmetric trilinear map $l_3:V_0\times V_0\times V_0\longrightarrow
V_1$ satisfying $l_3\circ\phi_0=\phi_1\circ l_3,$
   \end{itemize}
   such that for any $w,x,y,z\in V_0$ and $m,n\in V_1$, the following equalities are satisfied:
\begin{itemize}
\item[$\rm(a)$] $l_2(x,y)=-l_2(y,x),$
\item[$\rm(b)$] $l_2(x,m)=-l_2(m,x),$
\item[$\rm(c)$] $l_2(m,n)=0,$
\item[$\rm(d)$] $\dM l_2(x,m)=l_2(x,\dM m),$
\item[$\rm(e)$] $l_2(\dM m,n)=l_2(m,\dM n),$
\item[$\rm(f)$] $\phi_0(l_2(x,y))=l_2(\phi_0(x),\phi_0(y)),$
\item[$\rm(g)$]$\phi_1(l_2(x,m))=l_2(\phi_0(x),\phi_1(m)),$
\item[$\rm(h)$]$\dM l_3(x,y,z)=l_2(\phi_0(x),l_2(y,z))+l_2(\phi_0(y),l_2(z,x))+l_2(\phi_0(z),l_2(x,y)),$
\item[$\rm(i)$] $\dM l_3(x,y,\dM m)=l_2(\phi_0(x),l_2(y,m))+l_2(\phi_0(y),l_2(m,x))+l_2(\phi_1(m),l_2(x,y)),$
\item[$\rm(j)$] \begin{eqnarray*}
&&l_3(l_2(w,x),\phi_0(y),\phi_0(z))+l_2(l_3(w,x,z),\phi^2_0(y))\\
&&+
l_3(\phi_0(w),l_2(x,z),\phi_0(y))+l_3(l_2(w,z),\phi_0(x),\phi_0(y)) \\
&=&l_2(l_3(w,x,y),\phi^2_0(z))+l_3(l_2(w,y),\phi_0(x),\phi_0(z))+l_3(\phi_0(w),l_2(x,y),\phi_0(z))\\
&&+l_2(\phi^2_0(w),l_3(x,y,z))+l_2(l_3(w,y,z),\phi^2_0(x))+l_3(\phi_0(w),l_2(y,z),\phi_0(x)).
\end{eqnarray*}
   \end{itemize}
\end{defi}

We will denote a hom-Lie 2-algebra by
$(V_1\stackrel{\dM}{\longrightarrow}V_0,l_2,l_3,\phi_0,\phi_1)$.

\section{Hom-Lie algebras and $(\sigma,\tau)$-DGCAs}

It is well-known that a Lie algebra structure on a vector space $\frkk$ is equivalent to a differential graded commutative algebra structure on the exterior algebra $\Lambda\frkk^*$. In this section, we generalize this important fact to the hom-setting. First we introduce the notion of a $(\sigma,\tau)$-differential graded commutative algebra ($(\sigma,\tau)$-DGCA for short).

\begin{defi}
A $(\sigma,\tau)$-differential graded commutative algebra $(\huaA,\sigma,\tau,d_\huaA)$ consists of a graded commutative   algebra $\huaA=\oplus_k\huaA_k$, degree-$0$ algebra endomorphisms $\sigma$ and $\tau$, and a degree-$1$  operator $d_\huaA:\huaA_k\longrightarrow\huaA_{k+1}$, such that the following conditions are satisfied:
\begin{itemize}
  \item[\rm(1)] $d_\huaA^2=0;$
   \item[\rm(2)] $d_\huaA\circ \sigma=\sigma\circ d_\huaA,\quad d_\huaA\circ \tau=\tau\circ d_\huaA;$
    \item[\rm(3)] $d_\huaA(ab)=d_\huaA(a)\tau(b)+(-1)^{|a|}\sigma(a)d_\huaA(b),\quad \forall~a\in \huaA_{|a|},b\in\huaA_{|b|},$
\end{itemize}
where $|a|$ denotes the degree of the homogeneous element  $a\in\huaA$.
\end{defi}

For the vector space $\g$, let $\Lambda \g^*=\sum_k\wedge^k\g^*$ denotes the  exterior algebra of $\g^*$.
For any endomorphism $\alpha:\g\longrightarrow\g$, $\alpha^*:\g^*\longrightarrow\g^*$ denotes its dual map. Then, $\alpha^*$ naturally extends to an algebra morphism, for which we use the same notation, of $\Lambda \g^*$. More precisely, for any $\xi\in\wedge^k\g^*$, we have
$$
(\alpha^*\xi)(x_1,\cdots,x_k)=\xi(\alpha(x_1),\cdots,\alpha(x_k))
$$

Let
$(\frkg,[\cdot,\cdot],\alpha)$ be a hom-Lie algebra, define
a degree-1  operator
$d:\wedge^k\frkg^*\longrightarrow\wedge^{k+1}\frkg^*$ by
\begin{equation}\label{eq:operator d}
d\xi(x_1,\cdots,x_{k+1})=\sum_{i<j}(-1)^{i+j}\xi([x_i,x_j],\alpha(x_1),\cdots,\hat{x_i},\cdots,\hat{x_j},\cdots,\alpha(x_{k+1})),\quad \forall~\xi\in\wedge^k\frkg^*.
\end{equation}

\begin{pro}
  With the above notations, we have
  \begin{itemize}
    \item[\rm(i)] $d^2=0;$
     \item[\rm(ii)] $\alpha^*\circ d=d\circ\alpha^*;$
      \item[\rm(iii)] For all $\xi\in\wedge^k\g^*,~\eta\in\wedge^l\g^*$, the following equality holds:
      \begin{equation}\label{eq:derivation}
            d(\xi\wedge\eta)=d\xi\wedge\alpha^*\eta+(-1)^k\alpha^*\xi\wedge d\eta.
          \end{equation}
  \end{itemize}
  That is, $(\Lambda \g^*,\alpha^*,d)$ is an $(\alpha^*,\alpha^*)$-differential graded commutative algebra.
\end{pro}
\pf The fact that  $d^2=0$ follows from the hom-Jacobi identity. We omit details.

For $\xi\in\wedge^{k}\frkg^* $, by the equality
$\alpha([x_i,x_j])=[\alpha(x_i),\alpha(x_j)]$, we have
\begin{eqnarray*}
&&\alpha^*\circ d\xi(x_1,\cdots,x_{k+1})\\
&=&d\xi(\alpha(x_1),\cdots,\alpha(x_{k+1}))\\
&=&\sum_{i<j}(-1)^{i+j}\xi([\alpha(x_i),\alpha(x_j)],\alpha^2(x_1),\cdots,\hat{x_i},\cdots,\hat{x_j},\cdots,\alpha^2(x_{k+1}))\\
&=&\sum_{i<j}(-1)^{i+j}\alpha^*\xi([x_i,x_j],\alpha(x_1),\cdots,\hat{x_i},\cdots,\hat{x_j},\cdots,\alpha(x_{k+1}))\\
&=&d(\alpha^*\xi)(x_1,\cdots,x_{k+1}),
\end{eqnarray*}
which implies that (ii) holds.

To prove (iii),  first we let $k=1,\xi\wedge\eta\in\wedge^{1+l}\frkg^*$,
\begin{eqnarray*}
&&d(\xi\wedge\eta)(x_1,\cdots,x_{l+2})\\
&=&\sum_{i<j}(-1)^{i+j}\xi\wedge\eta([x_i,x_j],\alpha(x_1),\cdots,\hat{x_i},\cdots,\hat{x_j},\alpha(x_{l+2}))\\
&=&\sum_{i<j}(-1)^{i+j}\Big\{\xi([x_i,x_j])\eta(\alpha(x_1),\cdots,\hat{x_i},\cdots,\hat{x_j},\cdots,\alpha(x_{l+2}))\\
&&+\sum_{p<i}(-1)^p\xi(\alpha(x_p))\eta([x_i,x_j],\alpha(x_1),\hat{x_p},\cdots,\hat{x_i},\cdots,\hat{x_j},\cdots,\alpha(x_{l+2}))\\
&&+\sum_{i<p<j}(-1)^{p-1}\xi(\alpha(x_p))\eta([x_i,x_j],\alpha(x_1),\hat{x_i},\cdots,\hat{x_p},\cdots,\hat{x_j},\cdots,\alpha(x_{l+2}))\\
&&+\sum_{j<p}(-1)^{p-2}\xi(\alpha(x_p))\eta([x_i,x_j],\alpha(x_1),\hat{x_i},\cdots,\hat{x_j},\cdots,\hat{x_p},\cdots,\alpha(x_{l+2}))
 \Big\}\\
&=&d\xi\wedge\alpha^*\eta(x_1,\cdots,x_{l+2})-\alpha^*\xi\wedge
d\eta(x_1,\cdots,x_{l+2})
\end{eqnarray*}
 So, when $k=1$, we have
$$d(\xi\wedge\eta)=d\xi\wedge\alpha^*\eta+(-1)^1\alpha^*\xi\wedge d\eta.$$
Now, by induction on $k$. Assume that,  when $k=n$, we have
$$d(\xi\wedge\eta)=d\xi\wedge\alpha^*\eta+(-1)^n\alpha^*\xi\wedge
d\eta.$$
 For $ \omega\in\frkg^*$, then
$\xi\wedge\omega\in\wedge^{n+1}\frkg^*$, we have
\begin{eqnarray*}
d(\xi\wedge\omega\wedge\eta)
&=&d\xi\wedge\alpha^*(\omega\wedge\eta)+(-1)^n\alpha^*\xi\wedge
d(\omega\wedge\eta)\\
&=&d\xi\wedge(\alpha^*\omega\wedge\alpha^*\eta)+(-1)^n\alpha^*\xi\wedge(d\omega\wedge\alpha^*\eta+(-1)\alpha^*\omega\wedge
d\eta)\\
&=&(d\xi\wedge\alpha^*\omega+(-1)^n\alpha^*\xi\wedge
d\omega)\wedge\alpha^*\eta+(-1)^{n+1}(\alpha^*\xi\wedge\alpha^*\omega)\wedge
d\eta\\
&=&d(\xi\wedge\omega)\wedge\alpha^*\eta+(-1)^{n+1}\alpha^*(\xi\wedge\omega)\wedge
d\eta,
\end{eqnarray*}
which finishes the proof. \qed\vspace{3mm}

The converse of the above conclusion is also true. Thus, we have the following theorem, which generalize the classical result about the relation between Lie algebra structures and DGCA structures.

\begin{thm}\label{thm:dgca}
$(\frkg,[\cdot,\cdot],\alpha)$ is hom-Lie algebra if and only if
$(\Lambda \g^*,\alpha^*,d)$ is an $(\alpha^*,\alpha^*)$-differential graded commutative algebra.
\end{thm}
\pf
We only need to prove the sufficient condition. Assume that $(\Lambda \g^*,\alpha^*,d)$ is an $(\alpha^*,\alpha^*)$-differential graded commutative algebra.
First, define a skewsymmetric bracket operation $[\cdot,\cdot]:\wedge^2\g\longrightarrow\g$ on $\frkg$  as follow:
$$\langle\eta,[x_1,x_2]\rangle=-d\eta(x_1,x_2),  \quad\forall\eta\in\frkg^*,x_1,x_2\in\frkg.$$

 For $\eta\in\frkg^*,x_1,x_2\in\frkg$,  we have
\begin{eqnarray*}
d(\alpha^*\eta)(x_1,x_2)&=&-\langle\alpha^*\eta,[x_1,x_2]\rangle=-\langle\eta,\alpha([x_1,x_2])\rangle;\\
\alpha^*
d\eta(x_1,x_2)&=&-\langle\eta,[\alpha(x_1),\alpha(x_2)]\rangle.
\end{eqnarray*}
By $d\circ \alpha^*=\alpha^*\circ d$, we have
$$\alpha([x_1,x_2])=[\alpha(x_1),\alpha(x_2)],$$
which implies that $\alpha $ is an algebra endomorphism.

By \eqref{eq:derivation}, for $\xi,\eta\in\frkg^*,x_1,x_2,x_3\in\frkg$,  we have
\begin{eqnarray*}
d(\xi\wedge\eta)(x_1,x_2,x_3)&=&d\xi\wedge(\alpha^*\eta)(x_1,x_2,x_3)-(\alpha^*\xi)\wedge
d\eta(x_1,x_2,x_3)\\
&=&d\xi(x_1,x_2)\eta(\alpha(x_3))-d\xi(x_1,x_3)\eta(\alpha(x_2))+d\xi(x_2,x_3)\eta(\alpha(x_1))\\
&&-\xi(\alpha(x_1))d\eta(x_2,x_3)+\xi(\alpha(x_2))d\eta(x_1,x_3)-\xi(\alpha(x_3))d\eta(x_1,x_2)\\
&=&-\xi([x_1,x_2])\eta(\alpha(x_3))+\xi([x_1,x_3])\eta(\alpha(x_2))-\xi([x_2,x_3])\eta(\alpha(x_1)\\
&&+\xi(\alpha(x_1))\eta([x_2,x_3])-\xi(\alpha(x_2))\eta([x_1,x_3])
+\xi(\alpha(x_3))\eta([x_1,x_2])\\
&=&-\xi\wedge\eta([x_1,x_2],\alpha(x_3))+\xi\wedge\eta([x_1,x_3],\alpha(x_2))-\xi\wedge\eta([x_2,x_3],\alpha(x_1)).
\end{eqnarray*}
Thus, for all $\omega\in\wedge^2\g^*$, the following equality holds:
$$
d\omega(x_1,x_2,x_3)=-\omega([x_1,x_2],\alpha(x_3))+\omega([x_1,x_3],\alpha(x_2))-\omega([x_2,x_3],\alpha(x_1)).
$$
Now, for $ \xi\in\frkg^*$, $d\xi\in\wedge^2\frkg^*$. According to
$d\circ d=0$,  we have
\begin{eqnarray*}
0&=&d(d\xi)(x_1,x_2,x_3)\\
&=&-d\xi([x_1,x_2],\alpha(x_3))+d\xi([x_1,x_3],\alpha(x_2))-d\xi([x_2,x_3],\alpha(x_1))\\
&=&\xi([[x_1,x_2],\alpha(x_3)]+[[x_3,x_1],\alpha(x_2)]+[[x_2,x_3],\alpha(x_1)]),
\end{eqnarray*}
which implies that
$$[[x_1,x_2],\alpha(x_3)]+[[x_3,x_1],\alpha(x_2)]+[[x_2,x_3],\alpha(x_1)]=0.$$

Thus, $(\frkg,[\cdot,\cdot],\alpha)$ is a hom-Lie algebra.\qed

\section{Representations of hom-Lie algebras-revisited}

Let $V$ be a vector space, and $\beta\in GL(V)$. Define a
skew-symmetric bilinear bracket operation
$[\cdot,\cdot]_\beta:\frkg\frkl(V)\times\frkg\frkl(V)\longrightarrow\frkg\frkl(V)$
by
$$[A,B]_\beta=\beta A\beta^{-1}B\beta^{-1}-\beta B\beta^{-1}A\beta^{-1},\quad \forall~ A,B\in\frkg\frkl(V),$$
where $\beta^{-1}$ is the inverse of $\beta$. Denote by
$\Ad_\beta:\frkg\frkl(V)\longrightarrow\frkg\frkl(V)$ the adjoint action on $\gl(V)$, i.e.
$\Ad_\beta(A)=\beta A\beta^{-1}.$

\begin{pro}
With the above notations,
$(\frkg\frkl(V),[\cdot,\cdot]_\beta,\Ad_\beta)$ is a regular hom-Lie
algebra.
\end{pro}
\pf
Since $\Ad_\beta\circ \Ad_{\beta^{-1}}=\rm{id}$, it is obvious that $\Ad_\beta$ is invertible. Furthermore, we have
\begin{eqnarray*}
[\Ad_\beta(A),\Ad_\beta(B)]_\beta &=&[\beta A\beta^{-1},\beta
B\beta^{-1}]_\beta=\beta^2 A\beta^{-1}B\beta^{-1}\beta^{-1}-\beta^2
B\beta^{-1}A\beta^{-1}\beta^{-1}\\
&=& \Ad_\beta([A,B]_\beta),
\end{eqnarray*}
which implies that $\Ad_\beta$ is an
algebra
automorphism.

For all $A,B,C\in\frkg\frkl(V)$, we have
\begin{eqnarray*}
&&[[A,B]_\beta,\Ad_\beta(C)]_\beta+c.p.\\
&=&[\beta A\beta^{-1}B\beta^{-1},\beta C\beta^{-1}]_\beta-[\beta
B\beta^{-1}A\beta^{-1},\beta C\beta^{-1}]_\beta+c.p.\\
&=&\beta^2A\beta^{-1}B\beta^{-1}C({\beta^{-1}})^2-\beta^2C\beta^{-1}A\beta^{-1}B({\beta^{-1}})^2\\
&&-\beta^2B\beta^{-1}A\beta^{-1}C({\beta^{-1}})^2+\beta^2C\beta^{-1}B\beta^{-1}A({\beta^{-1}})^2+c.p.\\
&=&0.
\end{eqnarray*}
Thus, $(\frkg\frkl(V),[\cdot,\cdot]_\beta,\Ad_\beta)$ is a regular hom-Lie
algebra. \qed

\begin{thm}\label{thm:rephom}
Let $(\frkg,[\cdot,\cdot],\alpha)$ be a hom-Lie algebra, $V$ a
vector space, $\beta\in GL(V)$.
Then, $\rho:\frkg\longrightarrow\frkg\frkl(V)$ is a representation of $(\frkg,[\cdot,\cdot],\alpha)$
on   $V$ with respect to $\beta$ if and only if
$\rho:(\frkg,[\cdot,\cdot],\alpha)\longrightarrow(\frkg\frkl(V),[\cdot,\cdot]_\beta,\Ad_\beta)$
is a morphism of hom-Lie algebras.
\end{thm}
\pf
If $\rho :\frkg\longrightarrow\frkg\frkl(V)$ is a representation of
$(\frkg,[\cdot,\cdot],\alpha)$ on  $V$ with respect to $\beta$,  we have
\begin{eqnarray}
\label{eq:t1}\rho(\alpha(x))\circ\beta&=&\beta\circ\rho(x),\\
\label{eq:t2}\rho([x,y])\circ\beta&=&\rho(\alpha(x))\rho(y)-\rho(\alpha(y))\rho(x).
\end{eqnarray}
By \eqref{eq:t1}, we deduce that
$$
\rho\circ\alpha = \Ad_\beta\circ \rho.
$$
Furthermore, by \eqref{eq:t1} and \eqref{eq:t2}, we have
\begin{eqnarray*}
\rho([x,y])&=&\rho(\alpha(x))\beta\circ\beta^{-1}\rho(y)\beta^{-1}-\rho(\alpha(y))\beta\circ\beta^{-1}\rho(x)\beta^{-1}\\
&=&\beta\rho(x)\beta^{-1}\rho(y)\beta^{-1}-\beta\rho(y)\beta^{-1}\rho(x)\beta^{-1}\\
&=&[\rho(x),\rho(y)]_\beta.
\end{eqnarray*}
Thus, $\rho$ is morphism of hom-Lie algebras. The converse part is easy to be checked. The proof is completed.  \qed

\begin{cor}
 Let $(\frkg,[\cdot,\cdot],\alpha)$ be a regular hom-Lie algebra. Then, the adjoint representation $\ad:\g\longrightarrow\gl(\g)$, which is defined by $\ad_xy=[x,y]$, is a morphism from $(\frkg,[\cdot,\cdot],\alpha)$ to $(\gl(\g),[\cdot,\cdot]_\alpha,\Ad_\alpha)$.
\end{cor}

Let $(\frkg,[\cdot,\cdot],\alpha)$ be a hom-Lie algebra. Denote by
$C_\alpha^k(\frkg)=\{\xi\in\wedge^k\frkg^*|\alpha^*\xi=\xi\}$. Then, $(C^\bullet_\alpha(\g)=\oplus_kC_\alpha^k(\g),\wedge)$ is a subalgebra of the exterior algebra $(\Lambda\g^*,\wedge)$. It is straightforward to see that $(\oplus_kC_\alpha^k(\g),d)$ is a subcomplex of $(\Lambda\g^*,d)$. It is this subcomplex defined to be the cohomological complex of $\g$ in
\cite{homlie1}.

Let  $\rho:\frkg\longrightarrow\frkg\frkl(V)$ be a representation of
$(\frkg,[\cdot,\cdot],\alpha)$ on the vector space $V$ with respect to
$\beta\in\frkg\frkl(V)$.
The set of {\bf $k$-cochains} on $\frkg$ with values in $V$, which we
denote by $C^k(\frkg;V)$, is the set of skewsymmetric $k$-linear
maps from $\frkg\times\cdots\times\frkg$($k$-times) to $V$:
$$C^k(\frkg;V):=\{\varphi:\wedge^k\frkg\longrightarrow V ~\mbox{is a
linear map}\}.$$

Any $\beta\in\frkg\frkl(V)$ induces a map
$\bar{\beta}:C^k(\frkg;V)\longrightarrow C^k(\frkg;V)$ via
\begin{equation}
  \bar{\beta}(\varphi)(x_1,\cdots,x_k)=\beta\circ\varphi(x_1,\cdots,x_k),\quad \forall\varphi\in C^k(\frkg;V).
\end{equation}
$\alpha$ also induces a map
$\bar{\alpha}:C^k(\frkg;V)\longrightarrow C^k(\frkg;V)$ via
\begin{equation}
  \bar{\alpha}(\varphi)(x_1,\cdots,x_k)=\varphi(\alpha(x_1),\cdots,\alpha(x_k)),\quad \forall\varphi\in C^k(\frkg;V).
\end{equation}

A {\bf $k$-hom-cochain} on $\frkg$ with values in $V$ is defined to be a
$k$-cochain $\varphi\in C^k(\frkg;V)$ such that it is compatible
with $\alpha$ and $\beta$ is the sense that $\bar{\alpha}(\varphi)=\bar{\beta}(\varphi)$.
 Denote by $C_{\alpha,\beta}^k(\frkg;V)$ the set of $k$-hom-cochains on $\frkg$ with values in $V$.
 Obviously, if
$\varphi\in C_{\alpha,\beta}^k(\frkg;V)$, then both $\bar{\beta}(\varphi)$ and $
\bar{\alpha}(\varphi) $ are $k$-hom-cochains.

$C_{\alpha,\beta}^\bullet(\frkg;V)=\oplus_kC_{\alpha,\beta}^k(\frkg;V)$ is a $C^\bullet_\alpha(\g)$-module, where the action $\diamond:C_\alpha^l(\frkg)\times
C_{\alpha,\beta}^k(\frkg;V)\longrightarrow
C_{\alpha,\beta}^{k+l}(\frkg;V)$ is given by
 $$
 \eta\diamond\varphi(x_1,\cdots,x_{k+l})=\sum_\kappa \sgn(\kappa)\eta(x_{\kappa(1)},\cdots,x_{\kappa(l)})\varphi(x_{\kappa(l+1)},\cdots,x_{\kappa(k+l)}),
 $$
where $\eta\in C_\alpha^l(\frkg), \varphi\in
C_{\alpha,\beta}^k(\frkg;V),$ and the summation is taken over  $(l,k)$-unshuffles.


For $s=0,1,2,\ldots,$ define $d^s:C_{\alpha,\beta}^k(\frkg;V)\longrightarrow
C^{k+1}(\frkg;V)$ by
\begin{eqnarray*}
d^s\varphi(x_1,\cdots,x_{k+1})&=&\sum_{i=1}^{k+1}(-1)^{i+1}\rho(\alpha^{s+k}(x_i))\varphi(x_1,\cdots,\hat{x_i},\cdots,x_{k+1})\\
&&+\sum_{i<j}(-1)^{i+j}\varphi([x_i,x_j],\alpha(x_1),\cdots,\hat{x_i},\cdots,\hat{x_j},\cdots,\alpha(x_{k+1})).
\end{eqnarray*}

\begin{lem}
With the above notations, for all $\varphi\in
C_{\alpha,\beta}^k(\frkg;V)$,
$\bar{\alpha}(d^s\varphi)=\bar{\beta}(d^s\varphi).$
Thus, we obtain a well-defined map
$d^s:C_{\alpha,\beta}^k(\frkg;V)\longrightarrow
C_{\alpha,\beta}^{k+1}(\frkg;V)$.
\end{lem}

\pf
It follows from
\begin{eqnarray*}
&&d^s\varphi(\alpha(x_1),\cdots,\alpha(x_{k+1}))\\
&=&\sum_{i=1}^{k+1}(-1)^{i+1}\rho(\alpha^{s+k}\circ\alpha(x_i))\varphi(\alpha(x_1),\cdots,\hat{ x_i},\cdots,\alpha(x_{k+1}))\\
&&+\sum_{i<j}(-1)^{i+j}\varphi([\alpha(x_i),\alpha(x_j)],\alpha^2(x_1),\cdots,\hat{ x_i},\cdots,\hat{ x_j},\cdots,\alpha^2(x_{k+1}))\\
&=&\sum_{i=1}^{k+1}(-1)^{i+1}\rho(\alpha^{s+k+1}(x_i))\big(\beta\varphi(x_1,\cdots,\hat{x_i},\cdots,x_{k+1})\big)\\
&&+\sum_{i<j}(-1)^{i+j}\beta\varphi ([x_i,x_j],\alpha(x_1),\cdots,\hat{x_i},\cdots,\hat{x_j},\cdots,\alpha(x_{k+1}))\\
&=&\sum_{i=1}^{k+1}(-1)^{i+1}\beta\circ\rho(\alpha^{s+k}(x_i))\varphi(x_1,\cdots,\hat{x_i},\cdots,x_{k+1})\\
&&+\sum_{i<j}(-1)^{i+j}\beta\circ\varphi([x_i,x_j],\alpha(x_1),\cdots,\hat{x_i},\cdots,\hat{x_j},\cdots,\alpha(x_{k+1}))\\
&=&\beta\circ d^s\varphi(x_1,\cdots,x_{k+1}).\qed
\end{eqnarray*}

\begin{thm}\label{thm:rep}
  Let $(\g,[\cdot,\cdot],\alpha)$ be a hom-Lie algebra, and $\rho:\g\longrightarrow\gl(V)$  a representation of $\g$ on $V$ with respect to $\beta\in\gl(V)$. Then, for any $s=0,1,2,\ldots,$ we have
  \begin{itemize}
    \item[\rm(i)] The map $d^s$ is a coboundary operator, i.e. $d^s\circ d^s=0$;
     \item[\rm(ii)] $d^s$ and $\beta$ are compatible in the sense that $\beta\circ d^s=d^{s+1}\circ \bar{\beta};$
     \item[\rm(iii)] For $\eta\in C_\alpha^l(\frkg), \varphi\in
C_{\alpha,\beta}^k(\frkg;V)$, we have
\begin{equation}\label{eq:derivation1}
  d^s(\eta\diamond\varphi)=d \eta\diamond\bar{\alpha}(\varphi)+(-1)^l\eta\diamond d^{s+l}\varphi.
\end{equation}
  \end{itemize}
\end{thm}
\pf The proof of $(d^s)^2=0$ is similar as the one given in the Appendix in \cite{homlie1}. We omit details.

For all $\varphi\in C_{\alpha,\beta}^k(\frkg;V)$, by $\rho(\alpha(x_i))\circ
\beta=\beta\circ\rho(x_i)$,
 we have
\begin{eqnarray*}
\beta\circ d^s\varphi(x_1,\cdots,x_{k+1})
&=&\sum_{i=1}^{k+1}(-1)^{i+1}\beta\rho(\alpha^{s+k}(x_i))\varphi(x_1,\cdots,\hat{x_i},\cdots,x_{k+1})\\
&&+\sum_{i<j}(-1)^{i+j}\beta\varphi([x_i,x_j],\alpha(x_1),\cdots,\hat{x_i},\cdots,\hat{x_j},\cdots,\alpha(x_{k+1}))\\
&=&\sum_{i=1}^{k+1}(-1)^{i+1}\rho(\alpha^{s+k+1}(x_i))\beta\varphi(x_1,\cdots,\hat{x_i},\cdots,x_{k+1})\\
&&+\sum_{i<j}(-1)^{i+j}\beta\varphi([x_i,x_j],\alpha(x_1),\cdots,\hat{x_i},\cdots,\hat{x_j},\cdots,\alpha(x_{k+1}))\\
&=&d^{s+1}(\bar{\beta}(\varphi))(x_1,\cdots,x_{k+1}),
\end{eqnarray*}
which implies that $\beta\circ d^s=d^{s+1}\circ \bar{\beta}.$

To prove that \eqref{eq:derivation1} holds, first let $l=1$, then $\eta\diamond\varphi\in
C_{\alpha,\beta}^{k+1}(\frkg;V)$. We have
\begin{eqnarray*}
&&d^s(\eta\diamond\varphi)(x_1,\cdots,x_{k+2})\\
&=&\sum_{i=1}^{k+2}(-1)^{i+1}\rho(\alpha^{s+k+1}(x_i))\eta\diamond\varphi(x_1,\cdots,\hat{x_i},\cdots,x_{k+2})\\
&&+\sum_{i<j}(-1)^{i+j}\eta\diamond\varphi([x_i,x_j],\alpha(x_1),\cdots,\hat{x_i},\cdots,\hat{x_j},\alpha(x_{k+2}))\\
&=&\sum_{i<j}(-1)^{i+j}\eta([x_i,x_j])\varphi(\alpha(x_1),\cdots,\hat{x_i},\cdots,\hat{x_j},\cdots,\alpha(x_{k+2}))\\
&&+\sum_{p<i}(-1)^{i+p}\eta(x_p)\rho(\alpha^{s+k+1}(x_i))\varphi(x_1,\cdots,\hat{x_p},\cdots,\hat{x_i},\cdots,x_{k+2})\\
&&+\sum_{i<p}(-1)^{i+p+1}\eta(x_p)\rho(\alpha^{s+k+1}(x_i))\varphi(x_1,\cdots,\hat{x_i},\cdots,\hat{x_p},\cdots,x_{k+2})\\
&&+\sum_{p<i<j}(-1)^{p+i+j}\eta(\alpha(x_p))\varphi([x_i,x_j],\alpha(x_1),\cdots,\widehat{x_{p,i,j}},\cdots,\alpha(x_{k+2}))\\
&&+\sum_{i<p<j}(-1)^{p+i+j+1}\eta(\alpha(x_p))\varphi([x_i,x_j],\alpha(x_1),\cdots,\widehat{x_{i,p,j}},\cdots,\alpha(x_{k+2}))\\
&&+\sum_{i<j<p}(-1)^{p+i+j}\eta(\alpha(x_p))\varphi([x_i,x_j],\alpha(x_1),\cdots,\widehat{x_{i,j,p}},\cdots,\alpha(x_{k+2}))\\
&=&d \eta\diamond
\bar{\alpha}(\varphi)(x_1,\cdots,x_{k+2})+(-1)^1\eta\diamond
d^{s+1}\varphi(x_1,\cdots,x_{k+2}).
\end{eqnarray*}
Thus, when $l=1$, we have
$$d^s(\eta\diamond\varphi)=d \eta\diamond
\bar{\alpha}(\varphi)+(-1)^1\eta\diamond d^{s+1}\varphi.$$
 By induction on $l$,  assume that when $l=n$, we have
$$d^s(\eta\diamond\varphi)=d \eta\diamond\bar{\alpha}(\varphi)+(-1)^n\eta\diamond d^{s+n}\varphi.$$
For $\omega\in C_\alpha^1(\frkg)$, $\eta\wedge\omega\in
 C_\alpha^{n+1}(\frkg)$,   we have
\begin{eqnarray*}
d^s((\eta\wedge\omega)\diamond\varphi)&=&d^s(\eta\diamond(\omega\diamond\varphi))\\
&=&d \eta\diamond\bar{\alpha}(\omega\diamond\varphi)+(-1)^n\eta\diamond d^{s+n}(\omega\diamond\varphi)\\
&=&(d \eta\wedge\omega)\diamond\bar{\alpha}(\varphi)+(-1)^n\eta\diamond(d \omega\diamond\bar{\alpha}(\varphi)+(-1)\omega\diamond d^{s+n+1}\varphi)\\
&=&(d \eta\wedge\omega+(-1)^n\eta\wedge
d \omega)\diamond\bar{\alpha}(\varphi)+(-1)^{n+1}(\eta\wedge\omega)\diamond
d^{s+n+1}\varphi\\
&=&d (\eta\wedge\omega)\diamond\bar{\alpha}(\varphi)+(-1)^{n+1}(\eta\wedge\omega)\diamond
d^{s+n+1}\varphi.
\end{eqnarray*}
The proof is completed.\qed

\begin{rmk}
  If $\frkk$ is a Lie algebra, $\rho:\frkk\longrightarrow\gl(V)$ is a representation if and only if there is a degree-$1$ operator $D$ on $\Lambda\frkk^*\otimes V$ satisfying $D^2=0$, and
  $$
  D(\xi\wedge \eta\otimes u)=d_\frkk\xi\wedge\eta\otimes u+(-1)^k\xi \wedge D(\eta\otimes u),\quad \forall~\xi\in \wedge^k\frkk^*,~ \eta\in\wedge^l\frkk^*,~u\in V,
  $$
  where $d_\frkk:\wedge^k\g^*\longrightarrow \wedge^{k+1}\g^*$ is the coboundary operator associated to the trivial representation. However, for a hom-Lie algebra $(\g,[\cdot,\cdot],\alpha)$, we can only obtain Theorem \ref{thm:rep}. The converse is not true due to the restriction on the hom-cochains.
\end{rmk}

\section{Omni-hom-Lie algebras and hom-Lie 2-algebras}

\begin{defi}\label{defi1}
Let $V$ be a vector space and $\beta\in GL(V)$. An omni-hom-Lie algebra is a quadruple $(\frkg\frkl(V)\oplus
V,\delta_\beta,\{\cdot,\cdot\}_\beta,\langle\cdot,\cdot\rangle)$,
which we denote by $\frkg\frkl(V)\oplus_{\beta} V$ for short, where
  $\delta_\beta:\frkg\frkl(V)\oplus
V\longrightarrow\frkg\frkl(V)\oplus V$ is a linear map given by
$$\delta_\beta(A+u)=\Ad_\beta(A)+\beta(u),\quad \forall A+u\in\frkg\frkl(V)\oplus
V,$$
and
$\{\cdot,\cdot\}_\beta:(\frkg\frkl(V)\oplus
V)\otimes(\frkg\frkl(V)\oplus V)\longrightarrow\frkg\frkl(V)\oplus V$
is a linear map given by
$$\{A+u,B+v\}_\beta=[A,B]_\beta+A(v),\quad \forall A+u,B+v\in\frkg\frkl(V)\oplus
V,$$ and $\langle\cdot,\cdot\rangle:(\frkg\frkl(V)\oplus
V)\times(\frkg\frkl(V)\oplus V)\longrightarrow V$ is a symmetric
bilinear $V$-valued pairing given by
$$\langle A+u,B+v\rangle=\frac{1}{2}(A(v)+B(u)).$$
\end{defi}

Note that the bracket operation $\{\cdot,\cdot\}_\beta$ is not
skew-symmetric:
$$\{A+u,A+u\}_\beta=A(u)=\langle A+u,A+u\rangle.$$

\begin{rmk}
  When $\beta=\idd$, an omni-hom-Lie algebra $(\frkg\frkl(V)\oplus
V,\delta_\beta,\{\cdot,\cdot\}_\beta,\langle\cdot,\cdot\rangle)$ reduces to Weinstein's omni-Lie algebra, which is introduced in \cite{Weinomni} when study the linearization of Courant algebroids.
\end{rmk}

\begin{pro}\label{pro0}
With the above notations, we have
\begin{itemize}
\item[\rm{(a)}]$\delta_\beta$ is an algebra automorphism, i.e. $$\delta_\beta(\{A+u,B+v\}_\beta)=\{\delta_\beta(A+u),\delta_\beta(B+v)\}_\beta;$$

\item[\rm{(b)}] The bracket operation $\{\cdot,\cdot\}_\beta$ satisfies the hom-Leibniz rule:
$$\{\delta_\beta(A+u),\{B+v,C+w\}_\beta\}_\beta=\{\{A+u,B+v\}_\beta,\delta_\beta(C+w)\}_\beta+\{\delta_\beta(B+v),\{A+u,C+w\}_\beta\}_\beta.$$
\end{itemize}
Thus, $(\frkg\frkl(V)\oplus  V,\{\cdot,\cdot\}_\beta,\delta_\beta)$ is a hom-Leibniz algebra. Furthermore, $\delta_\beta$ and the pairing $\langle\cdot,\cdot\rangle$ are compatible in the sense that
$$\beta\langle
A+u,B+v\rangle=\langle\delta_\beta(A+u),\delta_\beta(B+v)\rangle.$$
\end{pro}
\pf Since $\Ad_\beta$ is an algebra automorphism, we have
\begin{eqnarray*}
\delta_\beta(\{A+u,B+v\}_\beta)&=&\delta_\beta([A,B]_\beta+A(v))
=\Ad_\beta([A,B]_\beta)+\beta A(v)\\
&=&[\Ad_\beta(A),\Ad_\beta(B)]_\beta+\beta A(v).
\end{eqnarray*}
On the other hand, we have
\begin{eqnarray*}
\{\delta_\beta(A+u),\delta_\beta(B+v)\}_\beta&=&\{\Ad_\beta(A)+\beta(u),\Ad_\beta(B)+\beta(v)\}_\beta\\
&=&[\Ad_\beta(A),\Ad_\beta(B)]_\beta+\Ad_\beta(A)\beta(v)\\
&=&[\Ad_\beta(A),\Ad_\beta(B)]_\beta+\beta A(v).
\end{eqnarray*}
Thus, $\delta_\beta$ is an algebra automorphism.

By straightforward computations, we have
\begin{eqnarray*}
\{\delta_\beta(A+u),\{B+v,C+w\}_\beta\}_\beta&=&\{\Ad_\beta(A)+\beta(u),[B,C]_\beta+B(w)\}_\beta\\
&=&[\Ad_\beta(A),[B,C]_\beta]_\beta+\Ad_\beta(A)B(w)\\
&=&[\Ad_\beta(A),[B,C]_\beta]_\beta+\beta A\beta^{-1}B(w),\\
\{\{A+u,B+v\}_\beta,\delta_\beta(C+w)\}_\beta&=&\{[A,B]_\beta+A(v),\Ad_\beta(C)+\beta(w)\}_\beta\\
&=&[[A,B]_\beta,\Ad_\beta(C)]_\beta+[A,B]_\beta\beta(w)\\
&=&[[A,B]_\beta,\Ad_\beta(C)]_\beta+\beta A\beta^{-1}B(w)-\beta
B\beta^{-1}A(w),\\
\{\delta_\beta(B+v),\{A+u,C+w\}_\beta\}_\beta&=&\{\Ad_\beta(B)+\beta(v),[A,C]_\beta+A(w)\}_\beta\\
&=&[\Ad_\beta(B),[A,C]_\beta]_\beta+\Ad_\beta(B)A(w)\\
&=&[\Ad_\beta(B),[A,C]_\beta]_\beta+\beta B\beta^{-1}A(w).
\end{eqnarray*}
Therefor, (b) follows from the fact that $(\frkg\frkl(V),[\cdot,\cdot]_\beta,\Ad_\beta)$ is
a hom-Lie algebra.

At last, we have
\begin{eqnarray*}
\langle\delta_\beta(A+u),\delta_\beta(B+v)\rangle&=&\langle
\Ad_\beta(A)+\beta(u),\Ad_\beta(B)+\beta(v)\rangle\\
&=&\frac{1}{2}(\Ad_\beta(A)\beta(v)+\Ad_\beta(B)\beta(u))\\
&=&\frac{1}{2}\beta(A(v)+B(u))\\
&=&\beta\langle A+u,B+v\rangle.
\end{eqnarray*}
 The proof is finished. \qed

\begin{rmk}
Actually, there are a series of algebraic structures satisfying properties in Proposition \ref{pro0}.
For any integer $q$, $\beta\in GL(V)$, consider the quadruple $(\frkg\frkl(V)\oplus
V,\delta_\beta,\{\cdot,\cdot\}_\beta^q,\langle\cdot,\cdot\rangle_\beta^q)$,
which we denote by $\frkg\frkl(V)\oplus_{\beta}^q V$ for short,
where   $\{\cdot,\cdot\}_\beta^q:(\frkg\frkl(V)\oplus
V)\otimes(\frkg\frkl(V)\oplus V)\longrightarrow\frkg\frkl(V)\oplus V$
is a linear map given by
$$\{A+u,B+v\}_\beta^q=[A,B]_\beta+\Ad_{\beta^{-q}}(A)(v), $$
and
$\langle\cdot,\cdot\rangle_\beta^q:(\frkg\frkl(V)\oplus
V)\times(\frkg\frkl(V)\oplus V)\longrightarrow V$ is a symmetric
bilinear $V$-valued pairing given by
$$\langle A+u,B+v\rangle_\beta^q=\frac{1}{2}(\Ad_{\beta^{-q}}(A)(v)+\Ad_{\beta^{-q}}(B)(u)).$$

When $q=0$, we obtain the omni-hom-Lie algebra given in Definition \ref{defi1}. One can deduce that properties in
Proposition \ref{pro0} are also correct for
$\frkg\frkl(V)\oplus_{\beta}^q V$ similarly. We omit details.
\end{rmk}

We introduce Dirac structures of omni-hom-Lie algebras. In general, Dirac structures have deep and widely applications in mathematical physics, see \cite{D} for more details.
\begin{defi}
A Dirac structure of the omni-hom-Lie algebra
$\frkg\frkl(V)\oplus_{\beta} V$ is a maximal isotropic subvector
space $L$, which is $\delta_\beta$-invariant and closed under the bracket operation
$\{\cdot,\cdot\}_\beta$.
\end{defi}

The following conclusion is obvious.
\begin{pro}\label{po1}
Let $L$ be a Dirac structure of the omni-hom-Lie algebra
$\frkg\frkl(V)\oplus_{\beta} V$, then
$(L,\{\cdot,\cdot\}_\beta,\delta_\beta)$ is a hom-Lie algebra.
\end{pro}

Let $V$ be a vector space and $F:V\times V\longrightarrow V$  a
bilinear map, $\beta\in GL(V)$. Denote by $\ad_F$ the induced map
from $V$ to $\frkg\frkl(V)$, i.e.
$$\ad_F(u)(v)=F(u,v),\quad \forall~ u,v\in V.$$
Denote by $\frkG_F\subset\frkg\frkl(V)\oplus V$  the graph of the map
$\ad_F$.
\begin{thm}\label{thm1}
The triple $(V,F,\beta)$ is a regular hom-Lie algebra  if and only if $\frkG_F$ is a  Dirac structure of
the omni-hom-Lie algebra $\frkg\frkl(V)\oplus_{\beta} V$.
\end{thm}
\pf
First we have
\begin{eqnarray*}
\langle \ad_F(u)+u,
\ad_F(v)+v\rangle&=&\frac{1}{2}(\ad_F(u)(v)+\ad_F(v)(u))\\
&=&\frac{1}{2}(F(u,v)+F(v,u)),
\end{eqnarray*}
which implies that $F$ is skew-symmetric is equivalent to that
$\frkG_F$ is isotropic. If $\frkG_F$ is isotropic, then it is
naturally maximal isotropic. In fact, assume that $B+v$ satisfies
$$\langle \ad_F(u)+u,
B+v\rangle=0,\quad \forall u\in V.$$
Then we have
$$0=\ad_F(u)(v)+B(u)=-\ad_F(v)(u)+B(u)=(B-\ad_F(v))(u)=0,$$
for all $u\in V$, which implies that $B=\ad_F(v)$. Thus, $\frkG_F$ is
maximal isotropic.

If $\frkG_F$ is a $\delta_\beta$-invariant subspace,
$\ad_F(\beta(u))=\Ad_\beta(\ad_F(u))$. Let it act on $\beta(v)$, we
have
$$F(\beta(u),\beta(v))=\beta\circ F(u,v),$$
which implies that $\beta$ is an algebra morphism. Therefore, $\beta$
is an algebra automorphism if and only if $\frkG_F$ is a $\delta_\beta$-invariant subspace.

At last, we have
$$\{\ad_F(u)+u,\ad_F(v)+v\}_\beta=[\ad_F(u),\ad_F(v)]_\beta+\ad_F(u)(v).$$
Thus, the condition that $\frkG_F$ is closed under the bracket
operation $\{\cdot,\cdot\}_\beta$ is equivalent to that
\begin{equation}\label{eq:ad-morphism}
  \ad_F(\ad_F(u)(v))=[\ad_F(u),\ad_F(v)]_\beta.
\end{equation}
Let it act on $\beta(w)$, we have
\begin{eqnarray*}
   \ad_F(\ad_F(u)(v))(\beta(w))&=&F(\ad_F(u)(v),\beta(w))=F(F(u,v),\beta(w)),\\
 ~[\ad_F(u),\ad_F(v)]_\beta(\beta(w))&=&\beta
\ad_F(u)\circ\beta^{-1}\circ \ad_F(v)\circ\beta^{-1}\beta(w)\\&&-\beta
\ad_F(v)\circ\beta^{-1}\circ \ad_F(u)\circ\beta^{-1}\beta(w)\\
&=&\beta\circ F(u,\beta^{-1}\circ F(v,w))-\beta\circ
F(v,\beta^{-1}\circ F(u,w))\\
&=&F(\beta(u),F(v,w))-F(\beta(v),F(u,w)).
\end{eqnarray*}
Thus, \eqref{eq:ad-morphism} is equivalent to the hom-Jacobi identity
$$F(\beta(u),F(v,w))+F(\beta(v),F(w,u))+F(\beta(w),F(u,v))=0.$$
The proof is finished. \qed

\begin{rmk}
  One can further prove that there is a one-to-one correspondence between general Dirac structures of
$\frkg\frkl(V)\oplus_\beta V$ and regular hom-Lie algebra structures
on subspaces of $V$. We omit details.
\end{rmk}

At the end of this section, we study properties of the skewsymmetrization of the bracket $\{\cdot,\cdot\}_\beta$.
Define a skewsymmetric bracket
$\Courant{\cdot,\cdot}:\wedge^2(\frkg\frkl(V)\oplus
V)\longrightarrow\frkg\frkl(V)\oplus V$ by
$$
\Courant{A+u,B+v}=\half (\{A+u,B+v\}_\beta-\{B+v,A+u\}_\beta)=[A,B]_\beta+\frac{1}{2}(A(v)-B(u)).
$$
Denote by $J:\wedge^3(\frkg\frkl(V)\oplus V)\longrightarrow V$  its hom-Jacobiator, i.e.
\begin{equation}
  J(A+u,B+v,C+w)=\Courant{\delta_\beta(A+u),\Courant{B+v,C+w}}+c.p..
\end{equation}

The hom-Jacobiator $J$ is not equal to $0$. By straightforward computations,  we have

\begin{lem}\label{lem:J}
For all $A+u,B+v,C+w\in\gl(V)\oplus V$, we have
 \begin{eqnarray*}
&&J(A+u,B+v,C+w)\\
&=&\frac{1}{4}(\beta C\beta^{-1}B(u)-\beta B\beta^{-1}C(u)+\beta
A\beta^{-1}C(v)-\beta C\beta^{-1}A(v)+\beta B\beta^{-1}A(w)-\beta
A\beta^{-1}B(w)).
\end{eqnarray*}
Consequently, we have
$$
J\circ \delta_\beta=\beta\circ J.
$$
\end{lem}

Now consider the 2-term complex $V\stackrel{\id}{\longrightarrow}{\gl(V)\oplus V}$, in which the degree-1 part is $V$ and the degree-0 part is $\gl(V)\oplus V$, and the differential is the inclusion $\id$. Define $l_2$ and $l_3$ by
\begin{equation} \label{eq:omni-2}
\left\{\begin{array}{rlll}
l_2(e_1,e_2)&=&\Courant{e_1,e_2},&\quad\mbox{for $e_1, e_2 \in \gl(V)\oplus V$},\\
l_2(e,w)&=&\Courant{e,\id(w)},&\quad\mbox{for $e \in \gl(V)\oplus V, ~w \in V$},\\
l_3(e_1,e_2,e_3)&=&J(e_1,e_2,e_3),&\quad\mbox{for $e_1, e_2, e_3
\in \gl(V) \oplus V$}.\end{array}\right.
\end{equation}

\begin{thm}\label{thm:homLie2}
With the above notations,
 $(V\stackrel{\id}{\longrightarrow}{\gl(V)\oplus V},l_2,l_3,\delta_\beta,\beta)$ is  a hom-Lie $2$-algebra, where $l_2$ and $l_3$ are given by \eqref{eq:omni-2}.
\end{thm}
\pf It is obvious that Conditions (a)-(e) hold naturally. Condition (f) follows from
\begin{eqnarray*}
\delta_\beta\Courant{A+u,B+v}&=&\half \delta_\beta(\{A+u,B+v\}_\beta-\{B+v,A+u\}_\beta)\\
&=&\half(\{\delta_\beta(A+u),\delta_\beta(B+v)\}_\beta-\{\delta_\beta(B+v),\delta_\beta(A+u)\}_\beta)\\
&=&\Courant{\delta_\beta(A+u),\delta_\beta(B+v)}.
\end{eqnarray*}

Condition (g) follows from
$$\beta(\Courant{A+u,v})=\Courant{\delta_\beta(A+u),\beta(v)}=\frac{1}{2}\beta A(v).$$

Conditions (h) and (i) follows directly from the definition of $l_3$.

At last, by Lemma \ref{lem:J} and tedious computations, we can show that Condition (j) also holds. Thus,  $(V\stackrel{\id}{\longrightarrow}{\gl(V)\oplus V},l_2,l_3,\delta_\beta,\beta)$ is  a hom-Lie $2$-algebra. \qed

\section*{Acknowledgement }
Research supported by NSFC (11471139) and NSF of Jilin Province (20140520054JH).

\end{document}